\documentclass[a4paper]{article}

\pdfoutput=1

\usepackage[margin=1in]{geometry} % full-width

\usepackage{amsmath}
\usepackage{amsthm}
\usepackage{appendix}
\usepackage{amssymb}
\usepackage{array}
\usepackage{pgfplots}
\usepackage{tabularx}
\usepackage[printwatermark]{xwatermark}
\usepackage{xcolor}
\usepackage{graphicx}
\graphicspath{ {./} }
\usetikzlibrary{shapes.multipart, calc}
\usepackage{lipsum}

\usepackage[utf8]{inputenc}
\usepackage{hyperref}
\hypersetup{
	unicode,
	pdfauthor={Chase Smith, Alex Rusnak},
	pdftitle={Dynamic Merkle B-tree with Efficient Proofs},
	pdfsubject={},
	pdfkeywords={},
	pdfproducer={LaTeX},
	pdfcreator={pdflatex}
}

\usepackage[sort&compress,numbers,square]{natbib}
\theoremstyle{plain}
\newtheorem{theorem}{Theorem}

\theoremstyle{definition}
\newtheorem{definition}[theorem]{Definition}

\newtheorem{remark}[theorem]{Remark}

\usepackage{graphicx, color}
\graphicspath{{./}}

\usepackage{algorithm, algpseudocode} % use algorithm and algorithmic for typesetting algorithms
\usepackage{mathrsfs} % for \mathscr command

\usepackage{lipsum}

\title{Dynamic Merkle B-tree with Efficient Proofs}
\author{Chase Smith* \\ \texttt{chase@proxima.one, chasesmith@berkeley.edu}
   \and Alex Rusnak \\ \texttt{alex@proxima.one, alexrusnak@berkeley.edu} }

\begin{document}
\begin{figure}[!ht]
\centering

	\maketitle
	
	\begin{abstract}
	 We propose and define a recursive Merkle structure with q-mercurial commitments, in order to create a concise B-Merkle tree. This Merkle B-Tree builds on previous work of q-ary Merkle trees which use concise, constant size, q-mercurial commitments for intranode proofs. Although these q-ary trees reduce the branching factor and height, they still have their heights based on the key length, and are forced to fixed heights. Instead of basing nodes on q-ary prefix trees, the B Merkle Tree incorporates concise intranode commitments within a self-balancing tree.  The tree is based on the ordering of elements, which requires extra information to determine element placement, but it enables significantly smaller proof sizes. This allows for much lower tree heights (directed by the order of elements, not the size of the key), and therefore creates smaller and more efficient proofs and operations. Additionally, the B Merkle Tree is defined with subset queries that feature similar communication costs to non-membership proofs. Our scheme has the potential to benefit outsourced database models, like blockchain, which use authenticated data structures and database indices to ensure immutability and integrity of data. We present potential applications in key-value stores, relational databases, and, in part, Merkle forests. \\ 
	 \noindent\textbf{Keywords:} Merkle, Authenticated data structure, accumulator
	\end{abstract}
	\end{figure}

	\tableofcontents
 
 \section{Introduction}
The value of accumulating and proving elements in a trustless manner is invaluable for ensuring tamper-proof data and has a variety of use cases in blockchain, certificate transparency, and outsourced data models.

There are a variety of different methods for attaining data authentication, with the most common being the utilization of Merkle trees.  Merkle trees authenticate data by recursively hashing the values of child nodes, this creates a distinct "root" each unique set of values. Merkle proofs of data involve providing all of the child hashes on each node within the path. This enables a reconstruction of the root itself. 

A major bottleneck for Merkle trees is the size of the proofs which increase logarithmically with the number of values, and the branching factor of the tree. Increasing the branching factor of the tree is a naive solution to speed, and performance, but this further increases the size of the proofs. 

While current methods for Merkle structures have intra-node proofs that are linear in size to the branching factor of the trie itself, there are q-ary Merkle prefix trees that provide constant sized intra-node proofs that provide constant size regardless of the branching factor. These intra-node proofs utilize mercurial commitment schemes that are positionally binding, but that are bounded in the number of elements that can be accumulated.  
Despite the benefit that is attained with these prefix structures, they are still implemented as tries, which means that their proof sizes are based on the size of the key, instead of the number of elements. 

There does exist a self-balancing Merkle tree that has been proposed which is based on the structure of the traditional AVL tree. This structure significantly reduces proof size because its height is based on the number of elements that it accumulates. 

In this work, we provide a hybridization of self-balancing trees and q-ary Merkle trees to provide a B$^+$ Merkle tree that maintains much shorter heights. This improves the computational efficiency of updating traditional q-ary Merkle trees, as well as providing efficient range, membership, and non-membership proofs.

\section{Preliminaries}
In what follows, all adversaries are probabilistic polynomial time (PPT) algorithms with respect to a security parameter $\kappa$ except if stated otherwise. We use the notation $e$: $\mathbb{G}$ $\times$ $\mathbb{G}$ $\rightarrow$ $\mathbb{G}_t$ to denote a 2$\kappa$ symmetric (type 1) bilinear pairing in groups of prime order p $\geq$ 2. The constructions of \cite{Kate2010ConstantSizeCT} can easily be modified to work with pairings of types 2 and 3 as well.

\begin{definition}{\textbf{Discrete Logarithm Assumption.}} \\
Given $g$ is a generator of $\mathbb{G}^*$, where $\mathbb{G}^*$ = $\mathbb{G}$ or $\mathbb{G}_t$, and $\alpha$  $\in$ $\mathbb{Z}_p^*$, for every adversary $\mathcal{A}_{DL}$, Pr[$\mathcal{A}_{DL}$($g$, $g^{\alpha}$) = a] = $\epsilon(k)$.
\end{definition}

\begin{definition}{\textbf{t-Polynomial Diffie-Hellman Assumption.}} \\
Let $\alpha$ $\in_R \mathbb{Z}_{p}^{*}$. Given as input a (t+1)-tuple ($g$, $g^{\alpha^1}$, ... , $g^{\alpha^t}$) $\in \mathbb{G}^{t+1}$, for every adversary $\mathbb{A}_{t-polyDH}$, the probability $Pr$[$\mathbb{A}_{t-polyDH}$($g$, $g^{\alpha^1}$, ... , $g^{\alpha^t}$) = $\langle$ $\phi(x)$, g$^{\phi(\alpha)}$ $\rangle$ ]  =  $\epsilon$(k) for any freely chosen $\phi(x)$ $\in$ $\mathbb{Z}_p$[$x$] such that $2^{\kappa}$ $>$ deg($\phi$) $>$ $t$. 
\end{definition}

\begin{definition}{\textbf{t-Strong Diffie-Hellman Assumption.}} \\
Let $\alpha$ $\in$ $\mathbb{Z}^*_p$. Given as input a (t+1)-tuple $\langle$ $g$, $g^{\alpha^1}$, ... , $g^{\alpha^t}$ $\rangle$ $\in\mathbb{G}^{t+1}$, for every adversary $\mathcal{A}_{t-SDH}$, the probability $Pr$[$\mathcal{A}_{t-SDH}$($g$, $g^{\alpha^1}$, ... , $g^{\alpha^t}$) = $\langle$ c, e($g$, $g$)$^{\frac{1}{\alpha + c}}$ $\rangle$ ] = $\epsilon$(k) for any value of c $\in$ $\mathbb{Z}^*_p$  $\backslash$ $\{-\alpha\}$.
\end{definition}

\begin{definition}{\textbf{t-Bilinear Strong Assumption.}} \\
Let $\alpha$ $\in$ $\mathbb{Z}^*_p$. Given as input a (t+1)-tuple $\langle$ $g$, $g^{a^1}$, ... , $g^{a^t}$ $\rangle$ $\in\mathbb{G}^{t+1}$, for every adversary $\mathcal{A}_{t-BDSH}$, the probability $Pr$[$\mathcal{A}_{t-BSDH}$($g$, $\langle g^{a^1}$, ... , $g^{a^t}\rangle$ = $\langle$ c, e($g$, $g$)$^{\frac{1}{\alpha + c}}$ $\rangle$ ] = $\epsilon$(k) for any value of c $\in$ $\mathbb{Z}^*_p$ $\backslash$ $\{-\alpha\}$.
\end{definition}
\begin{definition}{\textbf{Polynomial Commitment.}} \\
A polynomial commitment scheme consists of six algorithms: \textbf{Setup}, \textbf{Commit}, \textbf{Open}, \textbf{VerifyPoly}. \textbf{CreateWitness}, and \textbf{VerifyEval}.
\begin{quote}
\textbf{Setup}(1$^{\kappa}$, t) generates an appropriate algebraic structure $\mathcal{G}$ and a commitment public-private key pair $\langle$ PK, SK $\rangle$ to commit to a polynomial of degree $\leq$ t. For simplicity, we add $\mathcal{G}$ to the public key PK. \textbf{Setup} is run by a trusted or distributed authority. Note that SK is not required in the rest of the scheme. \\ \\
\textbf{Commit}(PK, $\phi$(x)) outputs a commitment $\mathcal{C}$ to a polynomial $\phi$(x) for the public key PK, and some associated decommitment information $d$. (In some constructions, $d$ is null.) \\ \\
\textbf{Open}(PK, $\mathcal{C}$, $\phi$(x), $d$) outputs the polynomial $\phi$(x) used while creating the commitment, with decommitment information $d$. \\ \\
\textbf{VerifyPoly}(PK, $\mathcal{C}$, $\phi$(x), $d$) verifies that $\mathcal{C}$ is a commitment to $\phi$(x), created with decommitment information $d$. If so the algorithm outputs 1, otherwise it outputs 0. \\ \\
\textbf{CreateWitness}(PK, $\phi$(x), $i$, $d$) outputs $\langle$ i, $\phi$(i), w$_i$ $\rangle$, where w$_i$ is a witness for the evaluation $\phi$(i) of $\phi$(x) at the index i and $d$ is the decommitment information. \\ \\
\textbf{VerifyEval}(PK, $\mathcal{C}$, i, $\phi$(i), $w_i$) verifies that $\phi$(i) is indeed the evaluation at the index i of the polynomial committed in $\mathcal{C}$. If so the algorithm outputs 1, otherwise it outputs 0.
\end{quote}
\end{definition}

\subsection{Related work}

There are a variety of different methods for maintaining data authentication. Recent updates in RSA accumulators have led to constant proof sizes regardless of the number of elements \cite{Boneh2018BatchingTF}
These accumulators provide a variety of batching and opening techniques, but suffer from issues with range proofs and the large sizes of public keys. This results in significant costs for membership and non-membership proofs (1.5kb). Other common techniques are based on the recursive hashing of tries, known as Merkle trees \cite{dahlberg2016efficient}. The proof sizes of these prefix trees depend on the height of the trie, and the branching factor of the nodes. 

Current methods have intranode proofs that are linear in size to the branching factor of the trie itself.  Other lesser-known prefix structures \cite{Libert2010ConciseMV, Kate2010ConstantSizeCT} have been given different forms of intranode proofs that provide constant size, and are used to create more efficient q-ary prefix trees. These intranode proofs utilize mercurial commitment schemes that are positionally binding, but that are bounded in the number of elements that can be accumulated.  

While the implementations of prefix structures have constant size intranode proofs, they are still implemented as tries, which means that their proof sizes are based on the size of the key, instead of the number of elements. 

There does exist a self-balancing Merkle tree \cite{reyzin2017improving} that has been proposed which is based on the structure of the binary AVL tree. This structure significantly reduces proof size because its height is based on the number of elements that it accumulates. Thus far, we do not believe that there exists a self-balancing tree that incorporates constant size intranode proofs for larger branching factors.

\subsection{Our Contributions}

We propose a dynamic, self-balancing Merkle tree that utilizes a polynomial  commitment scheme for intranode proofs. This enables our Merkle tree to have a significantly larger branching factor, while maintaining constant-size intranode proofs. 

Additionally, we base our Merkle tree on the $B^+$ tree which offers much more efficient proof sizes when compared to the typical prefix trees seen in Merkle tree implementations. Furthermore, we present range and non-membership proofs that are similar in size to traditional membership proofs.

\section{Dynamic  $B^+$ Merkle tree} 

The dynamic $B^+$ Merkle tree that we propose prove constant intranode proofs by utilizing a modified version of the polynomial commitment scheme seen in \cite{Kate2010ConstantSizeCT}. This scheme has a bounded number of elements that can be accumulated, but serves as the perfect foundation for constant sized proofs in trees with a high branching factor. Audit paths and commitments of nodes to the Merkle tree behavior normally. 

In the case of a changed child value, a node can update its hash naively by rehashing the values within its tree. Aside from the hashing of the nodes, the Merkle $B^+$ tree is structured, update, and queried in the same manner as the traditional $B^+$ tree.
\begin{figure}[!h]
\begin{center}
\includegraphics[width=8cm]{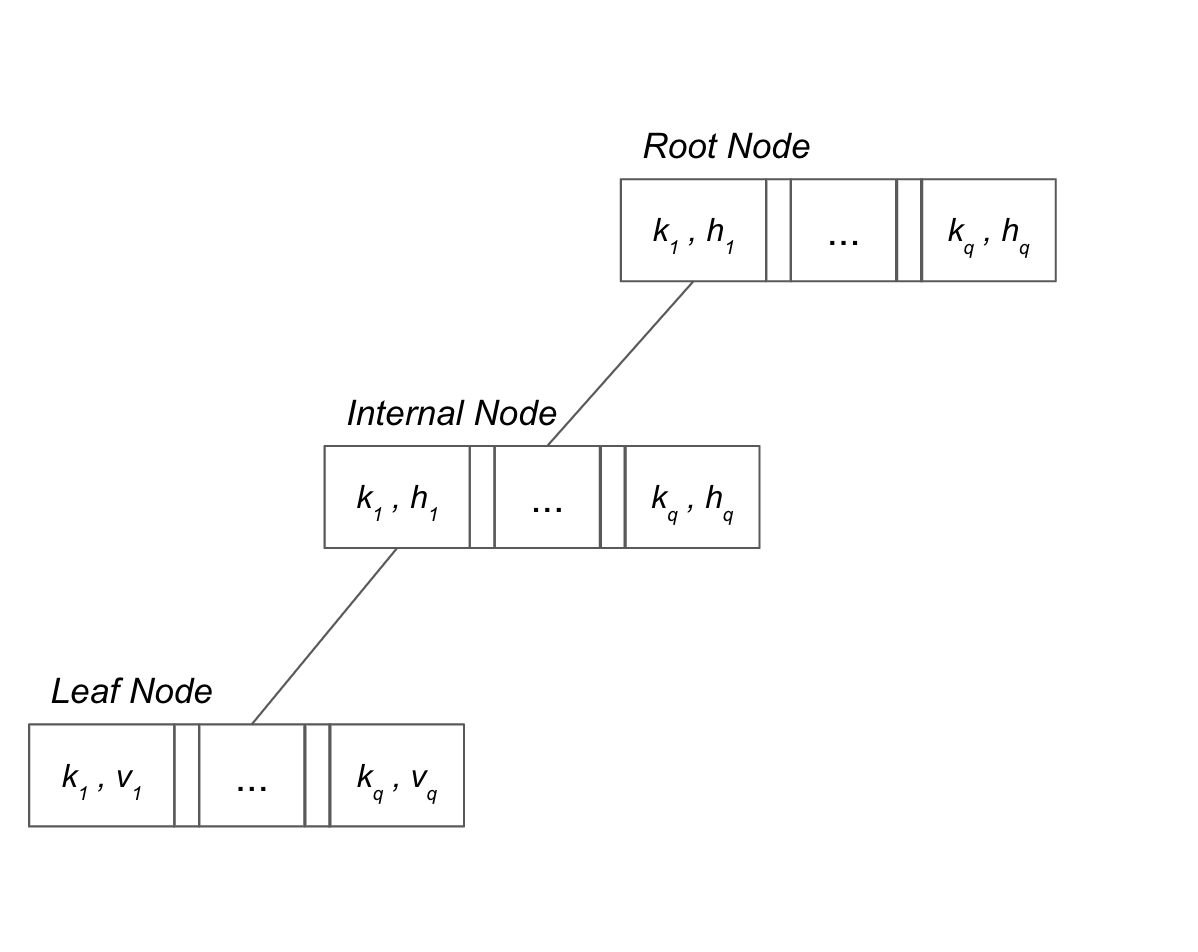}
\end{center}
 \caption{Dynamic Merkle $B^+$ tree}
\end{figure} \\
The scheme provides batched updates, constant sized batched witnesses, and efficient verification for key-value pairings within nodes. This enables the proofs to be verified using the commitment $\mathcal{C}$, the witness $W$ and the key-value pairs.
Additionally, 
\subsection{Node structure}

While the intranode hashes are based on the same input values, the each node is structured differently based on the type of the node itself [leaf, internal, root]. With ($k_i$, $h_i$) denoting the hash of the child node or value and ($k_i$, $pos_i$) denoting the position in the file system of the data itself. 
\subsubsection{Root Node}
  	\noindent\fbox{
    \parbox{\textwidth}{
    $hash_{node}$ \\
    $type_{node}$ \\ 
    $\phi(x)$ \\ 
    $\mathcal{C}_{}$ \\
    $S_{elements}$: [($k_1$, $h_1$), ..., ($k_b$, $h_b$)] \\
    $root_{previous}$
    }
}
\subsubsection{Internal Nodes}
  	\noindent\fbox{
    \parbox{\textwidth}{
    $hash_{node}$ \\
    $type_{node}$ \\ 
    $\phi(x)$ \\ 
    $\mathcal{C}_{}$ \\
    $S_{elements}$: [($k_1$, $h_1$), ..., ($k_b$, $h_b$)]
    }
}
\\
\subsubsection{Leaf Nodes}
  	\noindent\fbox{
    \parbox{\textwidth}{
    $hash_{node}$ \\
    $type_{node}$ \\ 
    $\phi(x)$ \\ 
    $\mathcal{C}_{}$ \\
    $S_{elements}$: [($k_1$, $h_1$), ..., ($k_b$, $h_b$)] \\
    $S_{positions}$: [($k_1$, $pos_1$), ..., ($k_b$, $pos_b$)]
    }
}
\\
\subsection{Node Commitments}

Each Merkle node commitment is comprised of the polynomial commitment ($\mathcal{C}$) and the type of the node. The polynomial commitment from \cite{Kate2010ConstantSizeCT} works to accumulate the key-value pairings of the child node hash, as well as their positions in the ordered elements within the node. 

In order to account for both key-value pair in an indexed position we transform the $k$ to $k'$ which incorporates the index, and denotes the sorted order of the key(s) within the elements of the node. This is important when calculating range and non-membership proofs. 
\\ \\ 
\noindent\fbox{
\parbox{\textwidth}{
\textbf{Node Commitment}
\\
$hash_{node}$ = $hash$($\mathcal{C}$ $||$ $type_{node}$)
\\ \\
\textbf{PolyCommitment} \\
$\mathcal{C}$ = $PolyCommit_{commit}$(PK, $S'$)
\\ \\ 
$S$  = [($k_1$, $v_1$), ..., ($k_b$, $v_b$)]\\ \\
$S'$ = [($k'_1$, $v_1$), ..., ($k'_b$, $v_b$)] \\ \\
$k_i'$ = $hash$ ($k_i$ $||$ $i$) \\ \\
\textbf{PolyVerification} \\ 
$PolyCommit_{verify}$ (PK, $\mathcal{C}$, $W_{i,j}$, [($k_i$, $v_i$), ..., ($k_j$, $v_j$)]) 
\\ \\
$W_{i, j}$ = witness for the elements $i$ through $j$
} }

\remark{} 
 The witness for a commitment denotes the value that is utilized by the Verifier to audit the data. This witness element remains constant regardless of the number of elements. Note that each commitment requires a polynomial interpolation of the keys and values such that $\phi(k_i)$ = $v_i$. This polynomial interpolation can be done in $q$ time with Horner's Method, where $q$ denotes the number of key-value pairings. There are further implementations of Horner's method that can be parallelized such that the complexity is $\frac{n}{k}$ with $k$ threads.

\subsection{Proofs}
The dynamic $B^+$ Merkle tree incorporates proofs for membership, as well as non-membership and range proofs. While proofs of range and non-membership require more elements that proofs of membership, they remain constant-sized. Though it should be noted that the proof of non-membership is equivalent to a proof of range.

\begin{quote}
\subsubsection{Membership}
The audit path for a proof of membership is comprised of the $B^+$ tree search path such that for every node the key that is selected is greater than or equal to the search $key$. 
\begin{figure}[h]
\begin{center}
\includegraphics[width=8cm]{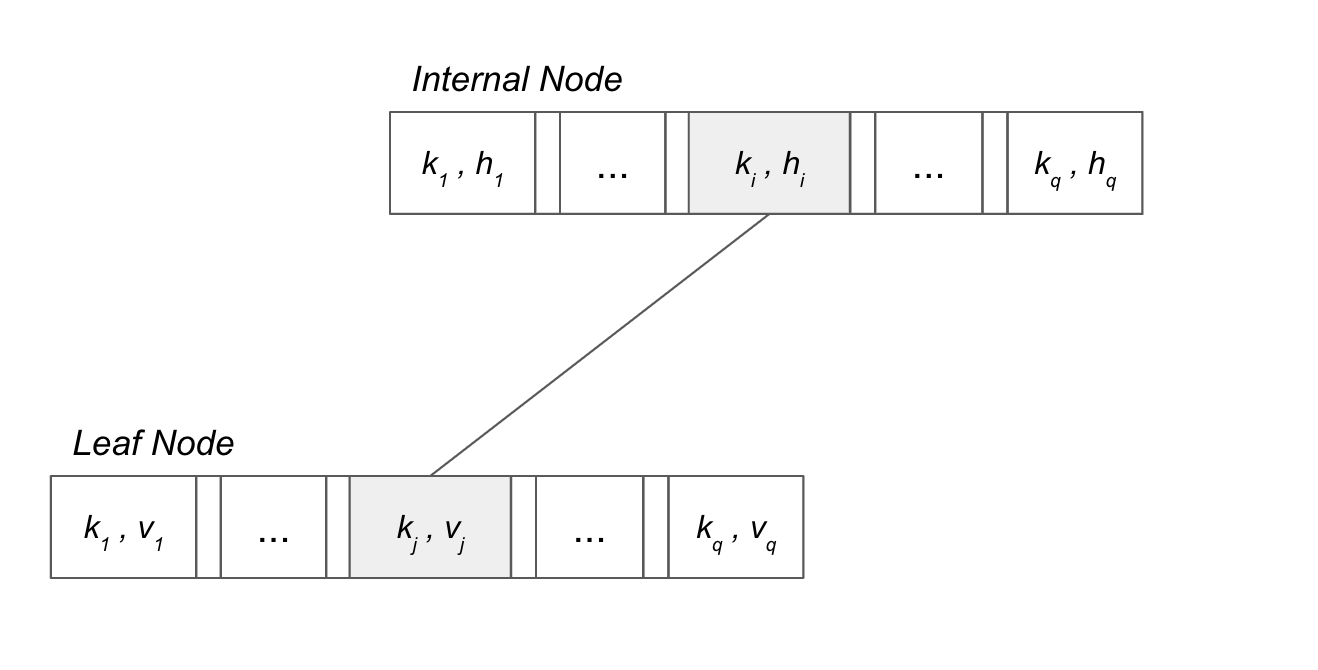}
\end{center}
 \caption{Dynamic Merkle $B^+$ tree}
\end{figure} \\
With this proof setup, each node in the audit path requires 3 group elements to be verified: the key $k_i$ to the key-value pairing, the witness $w_i$, and the index $i$ of the key. Note that the commitment $\mathcal{C}$ itself is not required because it is used as the hash value for the next node.
\end{quote}
\begin{quote}
\subsubsection{Non-membership}
Non-membership proofs comprise of two different cases for proving non-membership. Both audit paths utilize an extra element for a node hash, and adds one additional index and key to the complete proof. With this proof setup, all nodes along the audit path require 3 elements except for each interval node in the audit path which requires 4 group elements to be verified: the keys $k_i$ and $k_{i+1}$ to the key-value pairing, the witness $w_{i, i+1}$, and the index $i$ of the key. Note that the commitment $\mathcal{C}$ itself is not required because it is used as the hash value for the next node, and the witness is of constant size.
\\ \\ 
\textbf{Case 1} \\ 
When the search key is greater than $k_{i+1}$ and less than $k_i$ in some leaf node, then the proof is equivalent to the membership proof, except with the incorporation of two elements. In this case there is only one interval node: the leaf node itself. 
\\ \\ 
\textbf{Case 2} \\ 
When the search key is between two leaf nodes, then the internal node keys are required to verify the non-membership proof.  In this case there is one interval node: the internal node, but the audit path is split such there exists a range of keys with the smallest being smaller than the search key, and the greatest being larger than the search key. 
\\ \\ 
\begin{figure}[!h]
\begin{center}
\includegraphics[width=8cm]{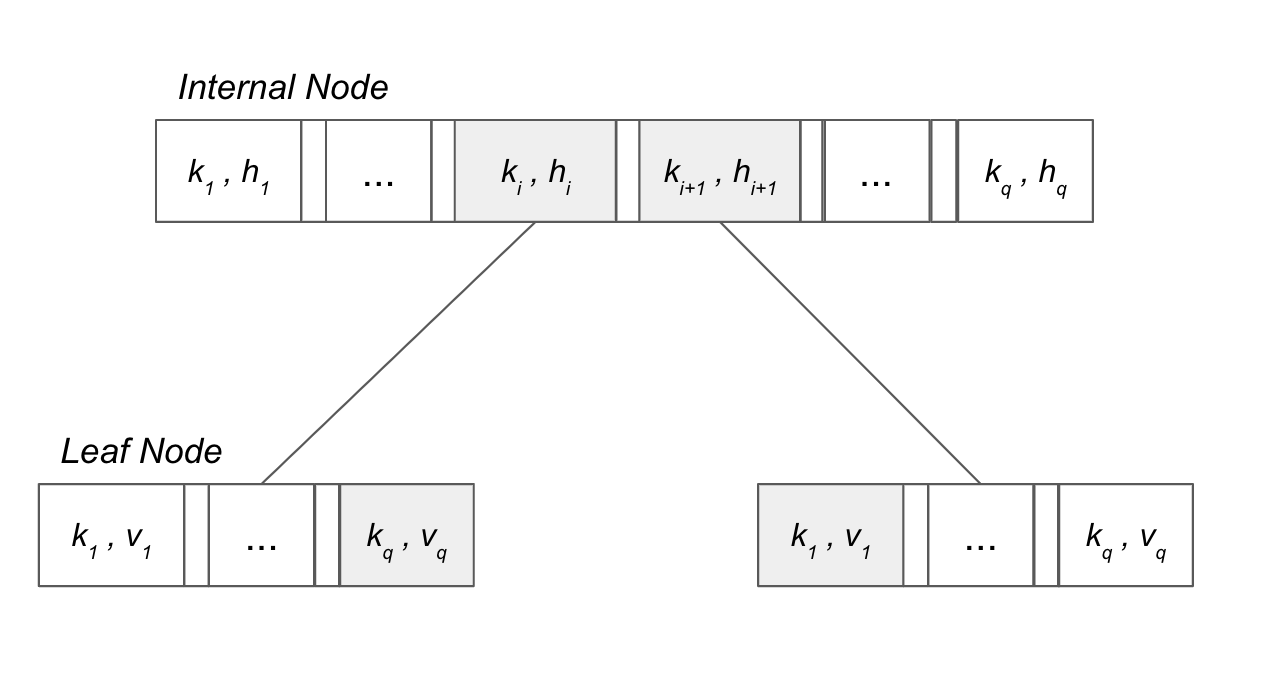}
\end{center}
 \caption{Dynamic Merkle $B^+$ tree}
\end{figure} \\
\end{quote}
\newpage
\begin{quote}
\subsubsection{Range}
The proof for the range proof incorporates two audit paths, and $log_qn$ $m$ bit vectors where $q$ is the branching factor, $m$ is the number of elements in the range and $n$ is the total number of elements in the tree. 
The two audit paths for the range proof capture the greatest and smallest elements in the range, or provide non-membership proofs for them. 
\\ 
\begin{figure}[!h]
\begin{center}
\includegraphics[width=8cm]{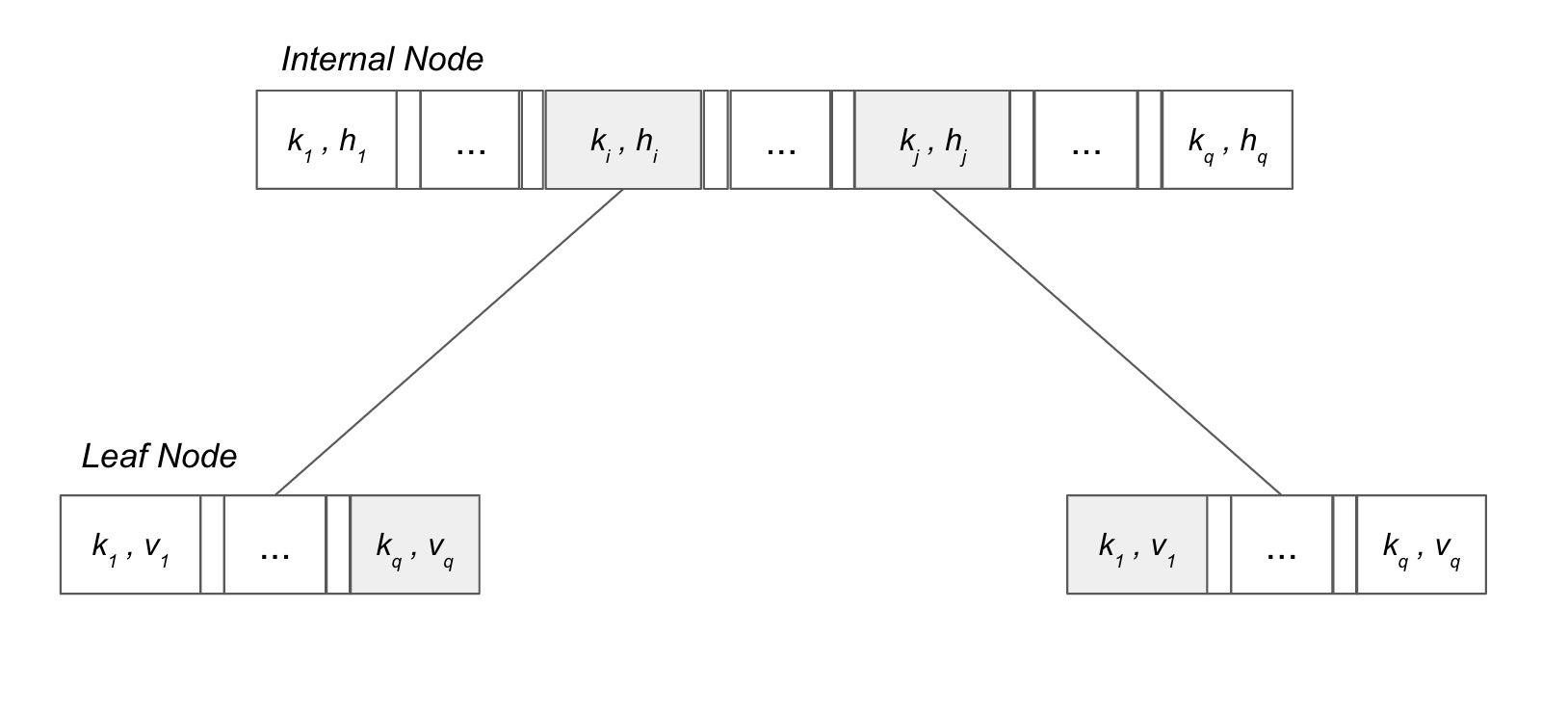}
\end{center}
 \caption{Dynamic Merkle $B^+$ tree}
\end{figure} \\
The bit vectors are also required for a range proof because the nodes within the $B^+$ can have between $\frac{q}{2}$ and $q-1$ elements. Thus, each bit vector is used to denote which keys are the endpoints for each node. Each bit vector has a size of $\frac{m}{q^j}$ with $j$ denoting the height of the Merkle tree. This adds $\Sigma^{h}_{j=0}\frac{m}{q^j}$ bits to the overall proof size. With these bit vectors and audit paths it is possible to build a complete Merkle proof.
\begin{figure}[!h]
\begin{center}
\includegraphics[width=8cm]{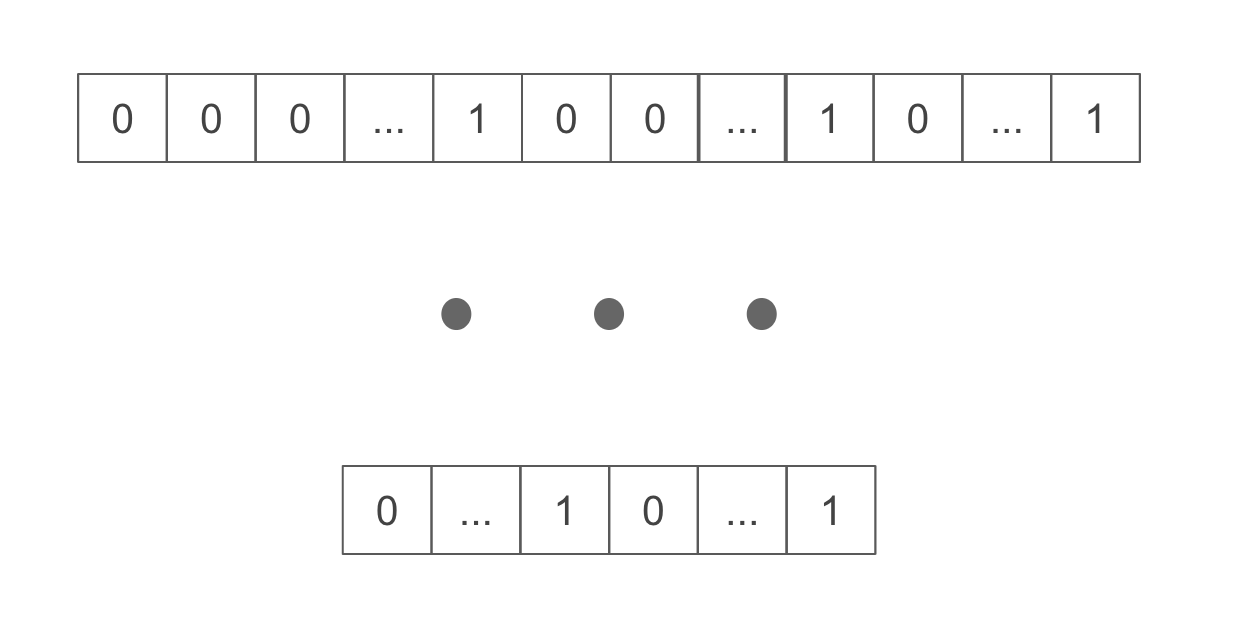}
\end{center}
 \caption{$B^+$  Merkle tree}
\end{figure}
\\
\end{quote}

\section{Complexity}

The computational and storage complexity of the dynamic $B^+$ Merkle tree is based on the height of the Merkle tree ($h$), the branching factor of the tree ($b$), and the number of elements ($n$). In this context, the height of the $B^+$ Merkle tree is logarithmic based on the branching factor and the number of elements. Additionally, the size required to store the indexing element for each key is defined by bytes$_{index}$. Furthermore, while the type of the node is required for the proof, it can be inferred by the placement of the node in the audit path.
 In order to verify commitments the Verifier must have a shared public key PK which has a size linear to the branching factor of the Merkle tree, but this shared public key does not change, and can be generated once and then re-used. 
 As stated previously, the type of node can be inferred from the audit path, and thus does not need to be considered in the proof itself. Likewise, proof sizes are proposed with the assumption that every query that is being audited, contains both the keys and the values.
\begin{table}[!h]
\caption{Computational Complexity and memory bounds of the dynamic B$^+$ Merkle tree operations.}
\begin{tabularx}{\textwidth} { 
  |>{\raggedright\arraybackslash}X 
  | >{\centering\arraybackslash}X 
| >{\raggedleft\arraybackslash}X |
  >{\raggedleft\arraybackslash}X | }
 \hline
 & Verification & Proof Size & Generation \\
 \hline
  \hline
 Non-membership & $log_q(n)\frac{q}{k}$  & $log_q(n)\frac{q}{k}$ & $log_q(n)log_2(q)$ \\
\hline
 Membership  & $log_q(n)\frac{q}{k}$  & $log_q(n)log_2(q)$ & $log_q(N)log_2(q)$ \\
\hline
 Range Proof  & $log_q(n)\frac{q}{k}$ + $m$  & $log_q(n)\frac{q}{k}$ + $m$ & $log_q(n)\frac{q}{k}$ + $m$ \\
\hline
\end{tabularx}
\end{table} 
We present a comparison between a variety of accumulator methods, as the number of elements increases. This is done with the assumption of 32 byte keys, and a Merkle tree branch out factor of 256 for both the q-ary prefix tree and the dynamic Merkle tree.
\\\\\\
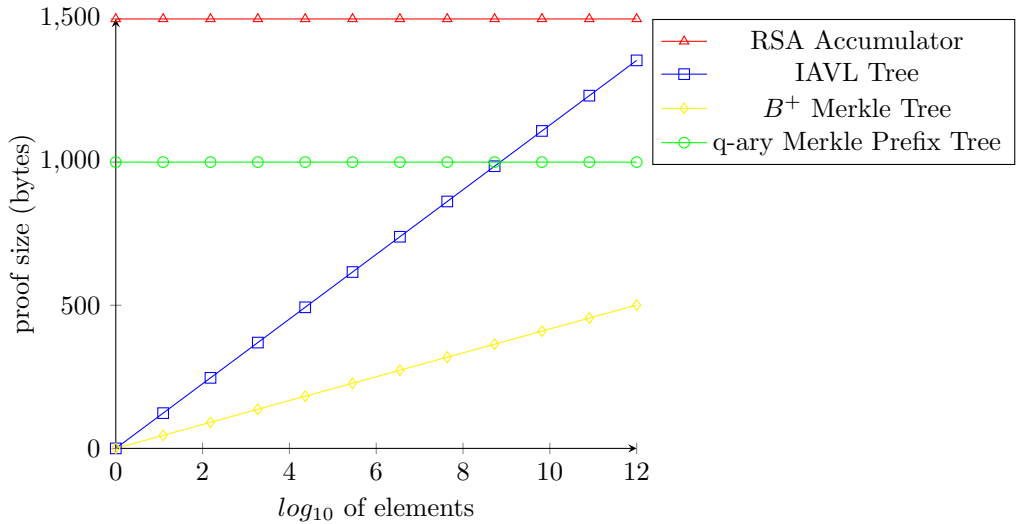
\begin{figure}[!h]
\begin{tikzpicture}
\begin{axis}[
    axis lines = left,
    xlabel = $log_{10}$ of elements,
    ylabel = proof size (bytes),
    legend pos = outer north east
]
%Below the red parabola is defined
\addplot [
    domain=0:12, 
    samples=12, 
    color=red,
    mark=triangle
]
{1500};
\addlegendentry{RSA Accumulator}
%Here the blue parabloa is defined
\addplot [
    domain=0:12, 
    samples=12, 
    color=blue,
    mark=square
    ]
    {log2(10^x) * (34)};
\addlegendentry{IAVL Tree}
 
\addplot [
    domain=0:12, 
    samples=12, 
    color=yellow,
    mark=diamond
    ]
    {(x/log10(200)) * (96)};
\addlegendentry{$B^+$ Merkle Tree}

 \addplot [
    domain=0:12, 
    samples=12, 
    color=green,
    mark=o
    ]
    {1000};
\addlegendentry{q-ary Merkle Prefix Tree}
\end{axis}
\end{tikzpicture}
\caption{Proof size comparison between B$^+$ Merkle tree and IAVL tree, q-ary tree, and RSA accumulator.}
\end{figure}

\remark{} It is important to note that both the q-ary tree and the RSA accumulator have a fixed proof size for membership proofs. This is a result of the q-ary Merkle Trie, as prefix tree, having a height based entirely the key size. The RSA accumulator the proof sizes are constant regardless of the number of elements. 

\section{Conclusion}

In conclusion, we have proposed a unique dynamic b-tree accumulation scheme that enables range proofs, and improves upon previous q-ary implements by having variable heights and modelling itself based on a dynamic, self-balancing tree. This scheme has promising applications in authenticated database systems. 

The main applications of such an efficient accumulation scheme, is nested in outsourced database models, as both a key-value store, and as a relational database index. 

Moreover, it has been made more effective by enabling batched updates, compression of proofs, incorporation of precomputation of witnesses, and pushing public string commitments and pairing-based computations to the client-side.

\section{Declarations}
\subsection{Funding}
The author received no financial support for the research, authorship, and/or publication of this article.

\subsection{Conflicts of interest}
The authors declare that they have no conflict of interest.

\bibliography{ms}

\begin{thebibliography}{5}
\providecommand{\natexlab}[1]{#1}
\providecommand{\url}[1]{\texttt{#1}}
\expandafter\ifx\csname urlstyle\endcsname\relax
  \providecommand{\doi}[1]{doi: #1}\else
  \providecommand{\doi}{doi: \begingroup \urlstyle{rm}\Url}\fi

\bibitem[Boneh et~al.(2018)Boneh, B{\"u}nz, and Fisch]{Boneh2018BatchingTF}
Dan Boneh, Benedikt B{\"u}nz, and Ben Fisch.
\newblock Batching techniques for accumulators with applications to iops and
  stateless blockchains.
\newblock \emph{IACR Cryptology ePrint Archive}, 2018:\penalty0 1188, 2018.

\bibitem[Dahlberg et~al.(2016)Dahlberg, Pulls, and
  Peeters]{dahlberg2016efficient}
Rasmus Dahlberg, Tobias Pulls, and Roel Peeters.
\newblock Efficient sparse merkle trees.
\newblock In \emph{Nordic Conference on Secure IT Systems}, pages 199--215.
  Springer, 2016.

\bibitem[Kate et~al.(2010)Kate, Zaverucha, and
  Goldberg]{Kate2010ConstantSizeCT}
Aniket Kate, Gregory~M. Zaverucha, and Ian Goldberg.
\newblock Constant-size commitments to polynomials and their applications.
\newblock In \emph{ASIACRYPT}, 2010.

\bibitem[Libert and Yung(2010)]{Libert2010ConciseMV}
Beno{\^i}t Libert and Moti Yung.
\newblock Concise mercurial vector commitments and independent zero-knowledge
  sets with short proofs.
\newblock In \emph{TCC}, 2010.

\bibitem[Reyzin et~al.(2017)Reyzin, Meshkov, Chepurnoy, and
  Ivanov]{reyzin2017improving}
Leonid Reyzin, Dmitry Meshkov, Alexander Chepurnoy, and Sasha Ivanov.
\newblock Improving authenticated dynamic dictionaries, with applications to
  cryptocurrencies.
\newblock In \emph{International Conference on Financial Cryptography and Data
  Security}, pages 376--392. Springer, 2017.

\end{thebibliography}
\newpage
\section{Appendix}
\subsection{B$^+$ tree}
Our dynamic $B^+$ Merkle tree utilizes the same structure as the traditional B$^+$ tree. Thus, the update, search, and deletions remain equivalent on an algorithmic level.  Note that batched operations can be done for both updates and deletions, and incur a cost logarithmic in the number of elements, and can be further optimized through parallelization. 
\begin{figure}[!h]
\begin{tabularx}{\textwidth} { 
  | >{\raggedright\arraybackslash}X 
  | >{\centering\arraybackslash}X | }
 \hline
  & Computation \\
 \hline
\hline
Search  & $log_qN$  \\
\hline
Range  & $log_qN$ + m  \\
\hline
Update  & $log_qN$  \\
\hline
Remove & $log_qN$  \\
\hline
\end{tabularx}
\caption{Computational complexity for $B^+$ tree operations, where $N$ is equal to the number of elements, $q$ is the branching factor of the B$^+$ tree, and $m$ is the number of elements within the range query.}
\end{figure}
\subsubsection{Searches}
The $B^+$ tree utilizes a search scheme similar to that of a binary tree. It recursively checks the keys indexed within the node, and follows the path of the first child that is greater than it. When the search reaches a leaf node, it checks for membership of the key and returns the value. If the key is not found in the leaf node, then the search is unsuccessful.  

\subsubsection{Updates}
Updates to the $B^+$ tree incorporate a search for the key in the $B^+$ tree, if the key is found, then the value in the tree is updated (in the case of the Merkle $B^+$ tree this involves the recomputing the hashes of the nodes effected. If the key is not found within a leaf node, then the key is inserted, and the nodes are updated to reflect the change. If an update causes a leaf node to overflow (the leaf reaches $q$ elements), then the $B^+$ splits the node via insertion as described below.   

\subsubsection{Insertions}
If a leaf node, internal node, or root node contain $q$ elements, then a new node must be allocated such that both nodes contain $\frac{q}{2}$ elements, and the middle element is added to the parent node. If the parent is full, split it too. Repeat until a parent is found that need not split. If the root splits, create a new root which has one key and two pointers. (That is, the value that gets pushed to the new root gets removed from the original node).
\subsubsection{Deletions}
Deletions operate the same way in insertions, with the key being found and removed the leaf node where it exists. If the key removal causes the internal, or leaf node to have less than $\frac{q}{2}$ elements, then the elements of its next youngest or oldest sibling are distributed to the node such that all nodes have at least $\frac{q}{2}$ elements. The hashes of the impacted nodes are then recomputed. 

\subsection{Polynomial Commitment}
The Polynomial commitment scheme that is utilized in the $B^+$ tree, and incorporates six different operations. 
\begin{quote}
\textbf{Setup}(1$^\kappa$, t). computes two groups $\mathbb{G}$, and $\mathbb{G}_t$ of prime order p (providing $\kappa-$bit security) such that there exists a symmetric bilinear pairing $e$ : $\mathbb{G}$ $\times$ $\mathbb{G}$ $\rightarrow$  $\mathbb{G}_t$ and for which the t-SDH assumption holds. We denote the generated bilinear pairing group as $\mathcal{G}$ = $\langle$ e, $\mathbb{G}$, $\mathbb{G}_t$ $\rangle$. Choose a generator g  $\in_R$ G. Let $\alpha$ $\in_R$ $\mathbb{Z^*_p}$ be SK, generated by a (possibly distributed) trusted authority. Setup also generates a (t + 1)-tuple $\langle$ g, g$^{\alpha}$, . . . , g$^{\alpha^t}$ $\rangle$ $\in$ $\mathbb{G}^{t+1}$ and outputs
PK = $\langle$ g, g$^{\alpha}$, . . . , g$^{\alpha^t}$ $\rangle$. SK is not required in the rest of the construction.
\\ \\
\textbf{Commit}(PK, $\phi$(x)) computes the commitment $\mathcal{C}$ = g$^{\phi(\alpha)}$ $\in$ $\mathbb{G}$ for polynomial $\phi$(x) $\in$ $\mathbb{Z}_p[X]$ of degree t or less. For $\phi$(x) = $\Sigma_{j=0}^{deg(\phi)}$ $\phi_j$x$^j$, it outputs $\mathcal{C}$ = $\Pi_{j=0}^{deg(\phi)}$ (g$^{\alpha^j}$)$\phi_j$ as the commitment to $\phi$(x). \\ \\
\textbf{Open}(PK, $\mathcal{C}$, $\phi$(x)) outputs the committed polynomial $\phi$(x).
\\ \\
\textbf{VerifyPoly}(PK, $\mathcal{C}$, $\phi$(x)) verifies that $\mathcal{C}$ $\stackrel{?}{=}$ g$^{\psi_i(\alpha)}$. If $\mathcal{C}$ = $\Pi_{j=0}^{deg(\phi)}$ (g$^{\alpha^j}$)$\phi_j$ for $\Sigma_{j=0}^{deg(\phi)}$ $\phi_j$x$^j$ the algorithm outputs 1, else it outputs 0. This only works when deg($\phi$)$\leq$ t. \\ \\
\textbf{CreateWitness}(PK, $\phi$(x), i) computes $\psi_i(x)$ = $\frac{\phi(x)-\phi(i)}{(x-i)}$ and outputs $\langle$ i, $\phi$(i), w$_i$ $\rangle$, where the witness $w_i$ = $g^{\psi_i(\alpha)}$ is computed in a manner similar to $\mathcal{C}$, above. \\ \\
\textbf{VerifyEval}(PK, $\mathcal{C}$, i, $\phi$(i), w$_i$) verifies that $\phi$(i) is the evaluation at the index i of the polynomial committed to by $\mathcal{C}$. If e($\mathcal{C}$, g) $\stackrel{?}{=}$ e(w$_i$, g$^\alpha$ /g$^i$) e(g, g)$^{\phi(i)}$, the algorithm outputs 1, else it outputs 0.
\end{quote}

\begin{figure}[!h]
\begin{tabularx}{\textwidth} { 
  | >{\raggedright\arraybackslash}X 
  | >{\centering\arraybackslash}X | }
 \hline
  & Computation \\
 \hline
 \hline
Update (batched)  & $\mathbb{O}$($log_q(N)$[$log_2(q)$ + $m$]   \\
\hline
Update  & $\mathbb{O}$($log_q(N)$  \\
\hline
\end{tabularx}
\\ \\ \\\\
\begin{tabularx}{\textwidth} { 
  |>{\raggedright\arraybackslash}X 
  | >{\centering\arraybackslash}X 
| >{\raggedleft\arraybackslash}X |
  >{\raggedleft\arraybackslash}X | }
 \hline
 & Verification & Proof Size & Generation \\
 \hline
  \hline
 Membership  & $\mathbb{O}$($log_q(N)$[$log_2(q)$)  & $\mathbb{O}$($log_q(N)$[$log_2(q)$ & $\mathbb{O}$($log_q(N)$ \\
\hline
 Membership (Batched)   & $\mathbb{O}$($log_q(N)$[$log_2(q)$  & $\mathbb{O}$($log_q(N)$[$log_2(q)$ & $\mathbb{O}$($log_q(N)$[$log_2(q)$  \\
\hline
\end{tabularx}
\caption{Computational complexity for polynomial operations}
\end{figure}
\end{document}